\documentclass[12pt]{article}
\usepackage[utf8]{inputenc}
\usepackage[a4paper, margin=1in]{geometry}
 \usepackage{setspace}
\usepackage{parskip}           
\usepackage{titlesec}
\usepackage{graphicx}      
\usepackage{float}

\usepackage{booktabs}
\usepackage{siunitx}

\usepackage{amsmath, amssymb}
\usepackage{newtxtext, newtxmath}

\usepackage{caption}

\usepackage{hyperref}
\usepackage{xcolor}
\usepackage{orcidlink}

\usepackage[numbers,compress]{natbib}

\newcommand{\mathsym}[1]{{}}
\newcommand{\unicode}[1]{{}}
 
\titleformat{\section}
  {\normalfont\bfseries\Large}{\thesection.}{1em}{}
 
\begin{document}
 
\captionsetup[figure]{labelfont=bf, textfont=normalfont, labelsep=period}
\captionsetup[table]{labelfont=bf, textfont=normalfont, labelsep=period}
 
\begin{center}
{\Large \textbf{Femtosecond Laser Induced Metallization in Silicon via Photon Momentum Mediated Band Transition}}\\[0.8cm]
{\large
\mbox{Ehatasham Haque}\orcidlink{0009-0002-4082-4349}$^{1, 2}$,
\mbox{Sunjana Tarannum}\orcidlink{0009-0009-7647-3991}$^{1}$, \mbox{Jannatul Shahrin Shoshi}\orcidlink{0009-0008-6174-2266}$^{1,2,3}$,
and \mbox{Mahdy Rahman Chowdhury}\orcidlink{0000-0003-3737-0315}$^{1,2,3}$\textsuperscript{*}
}\\[0.5cm]
{\small
$^{1}$Department of Electrical \& Computer Engineering,
North South University, Bashundhara, Dhaka, Bangladesh\\[0.2cm]
$^{2}$NSU Optics Lab, North South University, Bashundhara, Dhaka, Bangladesh\\[0.2cm]
$^{3}$NSU Center of Quantum Computing, North South University, Bashundhara, Dhaka, Bangladesh\\[0.4cm]
}
{\small \textsuperscript{*}Corresponding author:
\textcolor{blue}{\underline{\href{mailto:mahdy.chowdhury@northsouth.edu}{mahdy.chowdhury@northsouth.edu}}}}
\end{center}
\vspace{1cm}

\noindent \textbf{Abstract}\\[0.3cm]
Silicon's indirect bandgap fundamentally limits its optical
absorption, restricting its use in active photonic devices. A
two-stage mechanism for reversible transient metallization of
silicon is introduced: a nanoscale Au--Si picocavity relaxes the
momentum-conservation constraint through photon-momentum confinement, after which a femtosecond pulse injects hot carriers beyond the Mott density, collapsing the bandgap into a metallic state. The analysis predicts injected carrier densities reaching $1.42 \times 10^{21}$~cm$^{-3}$, which drives the Drude free-carrier response across the epsilon-near-zero point at approximately 1.76~\textmu m, so that the photoexcited region exhibits negative real permittivity and a collapsed field penetration depth throughout the near- and mid-infrared. Excitation is performed at 632.8~nm, where the photon energy exceeds the indirect gap and the momentum-relaxed absorption channel is active, while the metallic optical response is evaluated at longer probe wavelengths. Thermal analysis indicates that the lattice remains below silicon's melting threshold under the reported pulse conditions. This switchable behavior could enable ultrafast optical modulators, reconfigurable metasurfaces, and CMOS-compatible photonic components for monolithically integrated electro-optical systems.

\vspace{1cm}

\clearpage
\section{Introduction}
\label{intro}

Silicon has remained the dominant material platform in electronic
and photonic research for decades. This is primarily because of its mature fabrication infrastructure, favorable electronic properties, thermal stability and compatibility with large scale complementary metal oxide semiconductor (CMOS) integration \cite{ref1, ref2}. Nevertheless, its optical functionality is fundamentally constrained by its indirect bandgap. In crystalline silicon, the conduction band minimum and valence band maximum occur at different crystal momenta so an optical transition requires simultaneous conservation of energy and momentum and generally involves phonon assistance \cite{ref3,ref4,ref5} . This additional momentum requirement substantially reduces the optical transition probability compared with direct bandgap semiconductors and limits the efficiency of silicon for light absorption, emission and optoelectronic applications \cite{ref4,ref5}. Consequently, several approaches have been investigated to modify the electronic structure and optical response of silicon including strain engineering, electric field induced band structure modification, magnetic field effects and intense optical excitation \cite{ref6,ref7,ref8,ref9,ref10,ref11,ref12,ref13}. Although these approaches can substantially modify silicon's electronic or optical properties, achieving a controllable, localized and reversible transition from its conventional semiconducting state to a highly conducting metal like state remains challenging.

One promising route toward transient metallization is the generation of an extremely dense nonequilibrium electron-hole population through ultrafast optical excitation. When a sufficiently intense femtosecond pulse is incident on silicon, photons generate electron-hole pairs on a timescale much shorter than the characteristic timescale of lattice thermalization. As the photoexcited carrier density increases, carrier-carrier
interactions and Coulomb screening become progressively stronger. At low excitation densities, electrons and holes can form correlated excitonic states because of their mutual Coulomb attraction. However, with increasing carrier density  the electrostatic interaction between an electron and a hole is increasingly screened by the surrounding free carriers. The exciton binding energy consequently decreases  while the effective screening length becomes shorter. When the average carrier separation becomes comparable to the characteristic excitonic length scale the bound excitonic population is destabilized and the system enters the Mott transition regime. In photoexcited silicon, this density driven transition has been experimentally investigated using terahertz spectroscopy  where the exciton ionization ratio increases strongly with carrier density and a correlated metallic electron hole plasma emerges above the Mott density \cite{ref14,ref15}. Thus, the Mott criterion provides a physically meaningful density scale for identifying the transition from a dilute photoexcited semiconductor toward a dense  collectively responding carrier plasma.

The emergence of a dense electron hole plasma also changes the optical properties of silicon. Once a sufficiently large fraction of carriers behaves as mobile or weakly bound carriers  their collective electromagnetic response contributes strongly to
the complex dielectric function. Additionally, increasing the
carrier density directly increases the plasma frequency and progressively strengthens the free carrier contribution to the dielectric response. When the plasma frequency becomes sufficiently large relative to the excitation frequency, the real part of the dielectric function can become negative, producing a metal-like optical response. The imaginary component simultaneously reflects carrier damping and optical energy dissipation. Consequently, transient metallization should not be identified solely from an increase in absorption rather it is more rigorously associated with the simultaneous evolution of carrier density, complex permittivity and electromagnetic response. Established experimental measurements of silicon's dielectric response \cite{ref16} provide the baseline against which such excitation-dependent modifications of the transient optical state can be assessed.

Ultrafast excitation is particularly important because it can produce a high instantaneous carrier density while limiting the time available for substantial lattice heating during the initial electronic excitation. In the first stage following femtosecond excitation, the electronic system can therefore be driven far from equilibrium with rapid carrier generation, carrier-carrier scattering, and screening occurring before complete energy transfer to the lattice. Subsequent carrier phonon coupling transfers part of the absorbed energy into lattice degrees of freedom while carrier recombination and diffusion reduce the local carrier density. The metallic response is consequently expected to be transient it appears while the photoexcited carrier density remains above the relevant threshold and disappears as the carrier population relaxes below that threshold. This distinction is important because an electronically induced transient metallic state is fundamentally different from permanent structural metallization. At sufficiently high fluence, laser irradiation can instead produce excessive lattice heating, melting, ablation  or amorphization \cite{ref17,ref18,ref19}. Indeed, intense femtosecond excitation has been shown to generate dense electron hole plasmas and transient metal like optical responses in silicon \cite{ref17} while other studies have reported laser induced plasmonic coupling and structural modification under intense excitation \cite{ref17,ref18}. The key challenge is therefore to reach the high density electronic regime without converting the desired reversible electronic transition into irreversible thermal damage.

The density driven nature of metallization also distinguishes the present approach from simply increasing the incident laser intensity. The carrier density generated within a nanoscale interaction volume depends not only on the incident fluence but also on the local absorption coefficient, electromagnetic field enhancement, interaction volume, photon energy and pulse duration. In the present architecture, nanoscale photon confinement provides an established mechanism for increasing the optical coupling of silicon. Kharintsev et al. demonstrated that photon confinement below approximately 3 nm broadens the photon momentum distribution sufficiently to enable electronic transitions at the silicon band edge without phonon assistance resulting in substantially enhanced absorption \cite{ref20}. Related optical near field studies have also demonstrated momentum assisted coupling to indirect bandgap transitions in silicon \cite{ref21,ref22}. These studies establish the photon momentum mechanism as a foundation for increasing the optical absorption of silicon, however, they do not by themselves establish ultrafast Mott threshold crossing and transient metallization. In the present work, the momentum enhanced absorption is therefore used as a preconditioning mechanism that increases the number of photons absorbed within the nanoscale interaction region before the high fluence pulse drives the carrier population toward the metallic regime.

Here, we investigate a sequential two stage mechanism that combines established photon momentum enhanced absorption with ultrafast carrier density driven metallization. In first stage , a focused continuous wave(cw) optical field is confined within the Au-Si picocavity  producing nanoscale electromagnetic confinement and momentum enhanced optical transitions in the silicon region. In second stage, a high intensity femtosecond pulse is applied to this optically preconditioned region to generate a dense electron hole population. The resulting carrier density is then compared with the Mott transition threshold  while full wave  electromagnetic and thermal simulations are used to examine the resulting optical and thermal behavior. For the parameters investigated in this work, the calculated carrier density reaches (1.42 × $10^{21}$ cm$^{-3}$)  exceeding the Mott density range ($10^{18}$--$10^{20}$ cm$^{-3}$) adopted in the present analysis.The simulations further show enhanced optical absorption associated with the high carrier-density state and a transient metal-like response in the infrared, together with a thermal analysis indicating that the selected pulsed excitation can avoid sustained heating beyond the silicon melting regime. We emphasise at the outset that the pump photon energy and the spectral window of the metallic response are distinct: 632.8~nm excitation is used because it exceeds the indirect gap and couples efficiently through the momentum-relaxed channel, whereas the resulting free-carrier plasma renders the material metallic only below its plasma edge, which for the predicted carrier density lies near 1.76~\textmu while the thermal analysis indicates that the selected pulsed excitation can avoid sustained heating beyond the silicon melting regime. The central contribution of this work is therefore the proposed coupling of nanoscale photon momentum enhanced absorption with ultrafast high density carrier injection to establish a pathway toward reversible transient metallization of silicon. Unlike a permanent structural phase transformation, the proposed state is electronically driven and can  in principle  disappear as the carrier population recombines and relaxes below the metallization threshold. Such localized and dynamically controllable metal like silicon regions could provide opportunities for ultrafast optical switching, transient plasmonic structures, dynamically tunable absorbers, reconfigurable photonic elements and hybrid electronic photonic architectures subject to experimental validation of the predicted transient state.

 \begin{figure}[H]  
    \centering
    \includegraphics[width=0.8\textwidth]{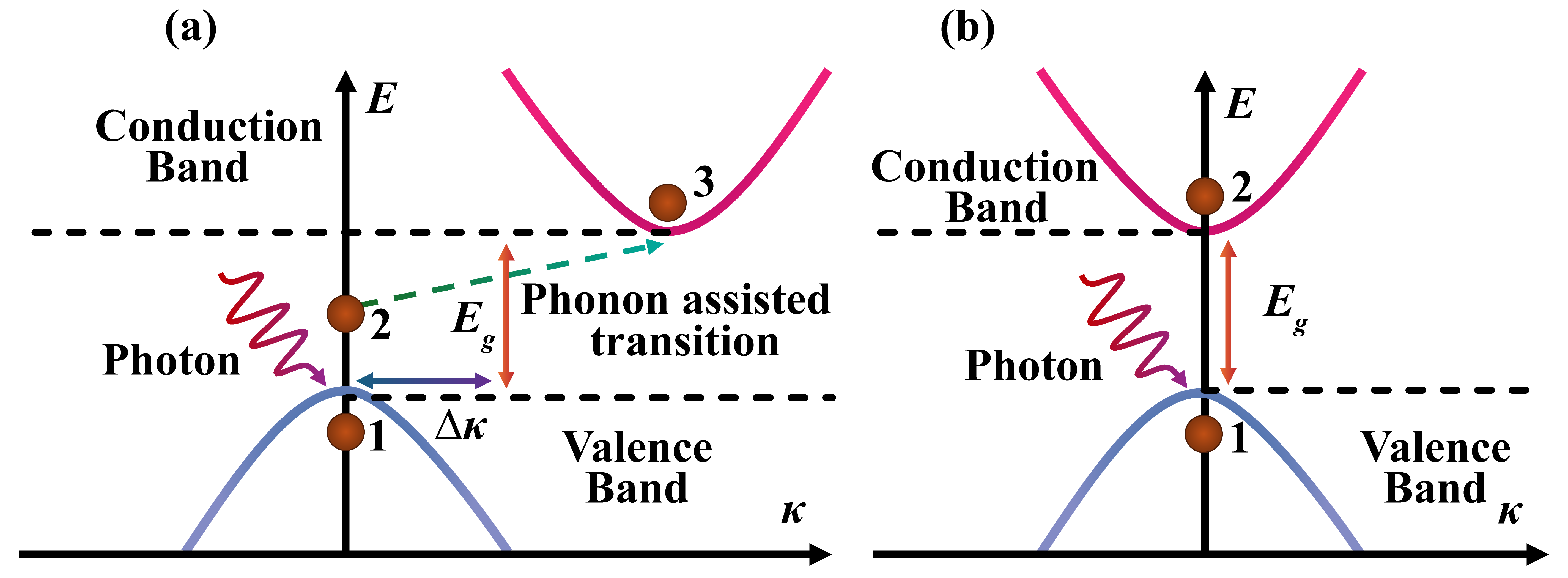}
    \caption{Categorization of Bandgaps. \textbf{(a)} Band diagram of an indirect bandgap material shown by electron-photon-phonon assisted transition. \textbf{(b)} Band diagram of a direct bandgap material.}
\end{figure}
 
\section{Setup}

\begin{figure}[H] 
    \centering
    \includegraphics[width=1\textwidth]{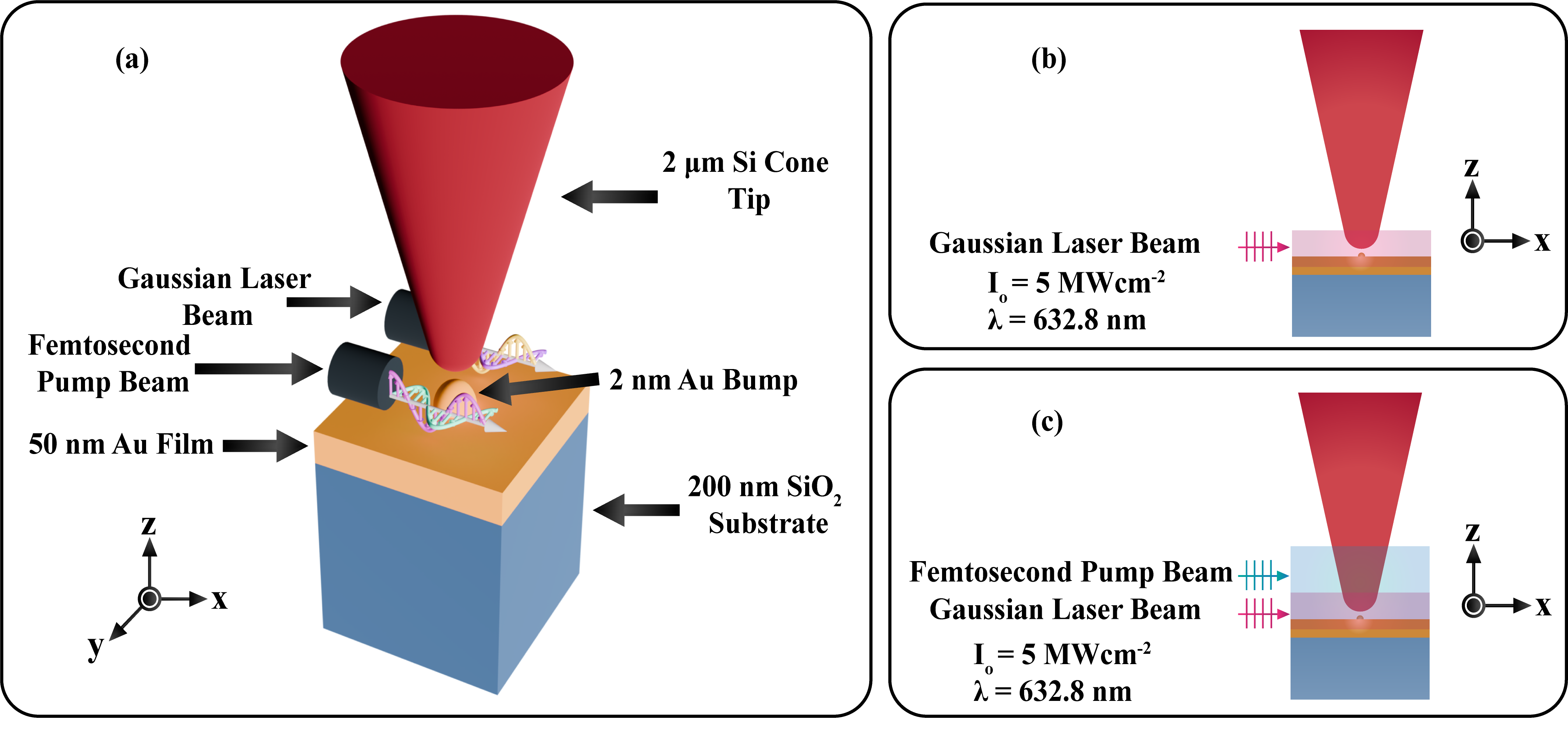}

   \caption{3D and 2D setup view for various types of optical excitation. (a) 3D geometric model showing dual excitation methods with a Gaussian laser beam and a femtosecond pump pulse (injected via a TFSF source formulation). (b) Setup for unexcited (indirect-bandgap) semiconducting conditions. (c) Setup for metallization conditions.}
\label{fig:2}
\end{figure}

The proposed three-dimensional architecture for semiconductor metallization is illustrated in Fig.~\ref{fig:2}. Additionally, Figs.~\ref{fig:2}b and \ref{fig:2}c depict photon momentum mediated band transitions and metallization setup modes, respectively. All optical and thermal analysis are performed employing the readily available Ansys Lumerical software suite. Particularly, all optical analysis are characterized by the Lumerical FDTD 2020 module and all thermal analysis are characterized by the Lumerical HEAT 2020 module of the software suite. Furthermore, all mathematical models and graphs are designed in the commercially available MATLAB and, open-source Python respectively. 

For optical analysis a rounded tip, cone-shaped Si tip under cw focused illumination was performed by using an Ansys/Lumerical FDTD solver. A mesh overlayer of 0.1~$\mathrm{nm}$ was utilized around the Au bump and the Si tip apex and a rougher 1~$\mathrm{nm}$ mesh for the rest of the structure.The 0.1~nm mesh overlayer was selected following a convergence study in which the mesh size around the Au bump and Si tip apex was progressively refined from 0.5~nm to 0.05~nm. The computed peak field enhancement factor $g$ and the Gaussian-beam absorbed power $P_{\mathrm{Gauss}}$ varied by less than 5\% between the 0.1~nm and 0.05~nm mesh cases, confirming that the reported results are converged with respect to spatial discretization (see Appendix~B for the convergence data).The Au bump radius ($\Delta r = 2$~nm) approaches the length scale at which nonlocal dielectric response and quantum tunneling across the junction can, in principle, modify the local-response predictions of classical FDTD. These corrections are most severe for symmetric metal--metal gaps below $\sim$0.5~nm, where electron tunneling directly bridges the junction. In the present Au--Si architecture, the junction is a metal--semiconductor interface rather than a metal--metal gap, and the relevant field confinement occurs predominantly on the silicon side of the junction, where the free carrier density prior to excitation is orders of magnitude lower than in gold. We therefore expect nonlocal corrections to primarily renormalize the peak field enhancement factor $g$ rather than qualitatively alter the carrier-generation mechanism reported here. A fully nonlocal or quantum-corrected treatment of the picocavity response is left for future work. The optical properties of Si and Au were imported from the Ansys/Lumerical material database. Particularly, for $\text{Si}$ tip and $\text{SiO}_2$ glass substrate their respective Palik models were chosen. For the $\text{Au}$ bump and $\text{Au}$ film, the Johnson and Christy model was chosen for its accurate representation of nanoscopic light-matter interactions. The Si tip apex was exposed to a 632.8~nm focused Gaussian beam with intensity of 5~MW/cm$^2$ and then excited sequentially with a 1~GW/cm$^2$ intensity femtosecond pulsed beam (injected via a TFSF source formulation) of 1000~fs pulse width. Additional parameters are provided in the appendix~\ref{app:B}. 

\section{Results and Analysis}
\label{results_discussion}
This section elaborates on our work with graphical proofs and explanations behind our notion, along with a mathematical basis behind the work. 

\subsection{Optical Signatures of Silicon Metallization}
\begin{figure}[H] 
    \centering
    \includegraphics[width=1\textwidth]{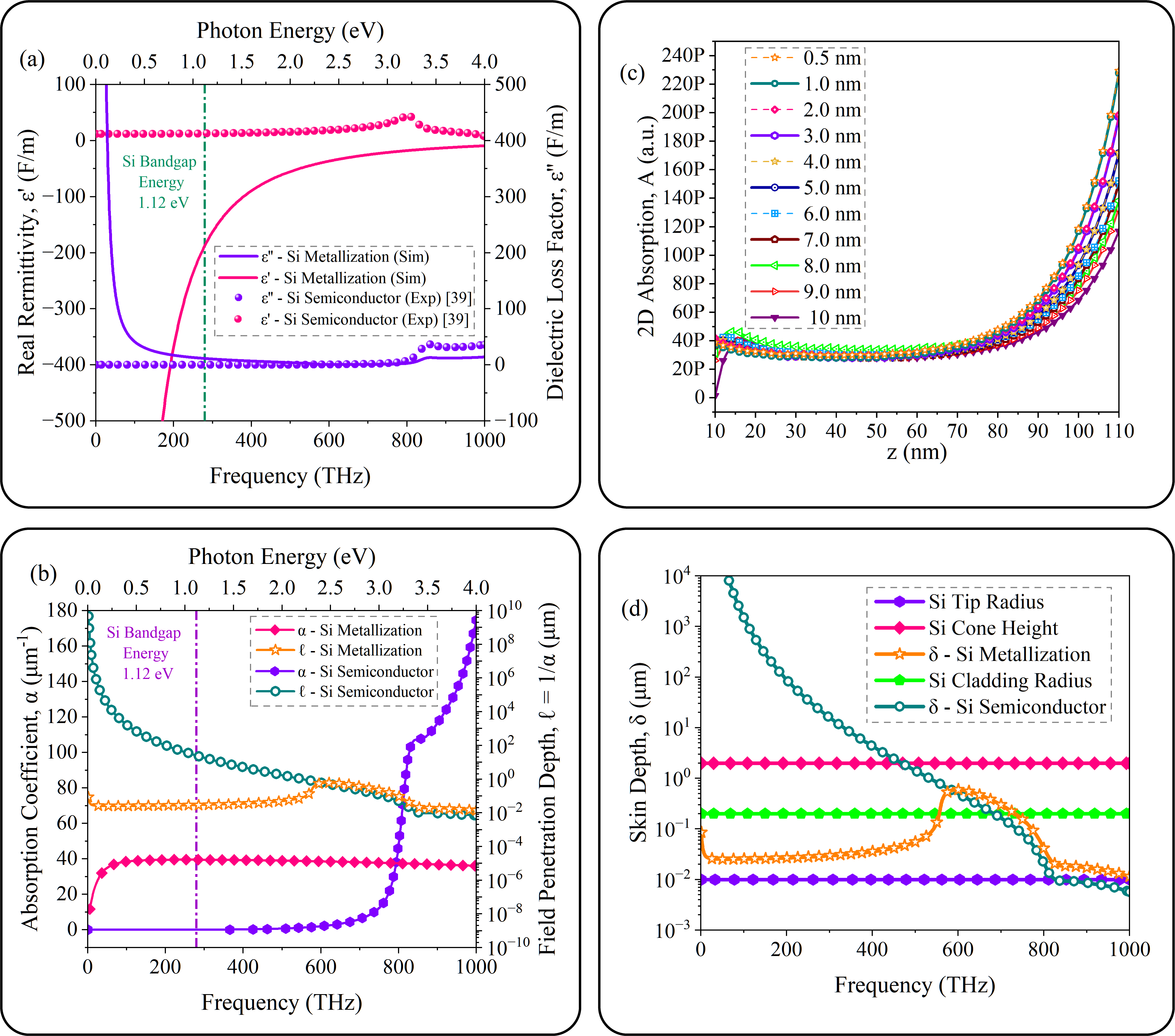}
    \caption{Computed optical properties of silicon semiconductor phase and metal phase. \textbf{(a)} Comparative view of permittivity and dielectric loss factor of silicon metallization and Palik silicon semiconductor. \textbf{(b)} Absorption coefficient and field penetration Depth for metallic and semiconductor phase of silicon. \textbf{(c)} Integrated energy absorbed by the Si tip along the z-direction as depicted in Fig.~\ref{fig:2}. \textbf{(d)} Skin depth as a function of frequency for silicon semiconductor and metal phase.}
    \label{fig:3}
\end{figure}

\begin{figure}[H] 
    \centering
    \includegraphics[width=1\textwidth]{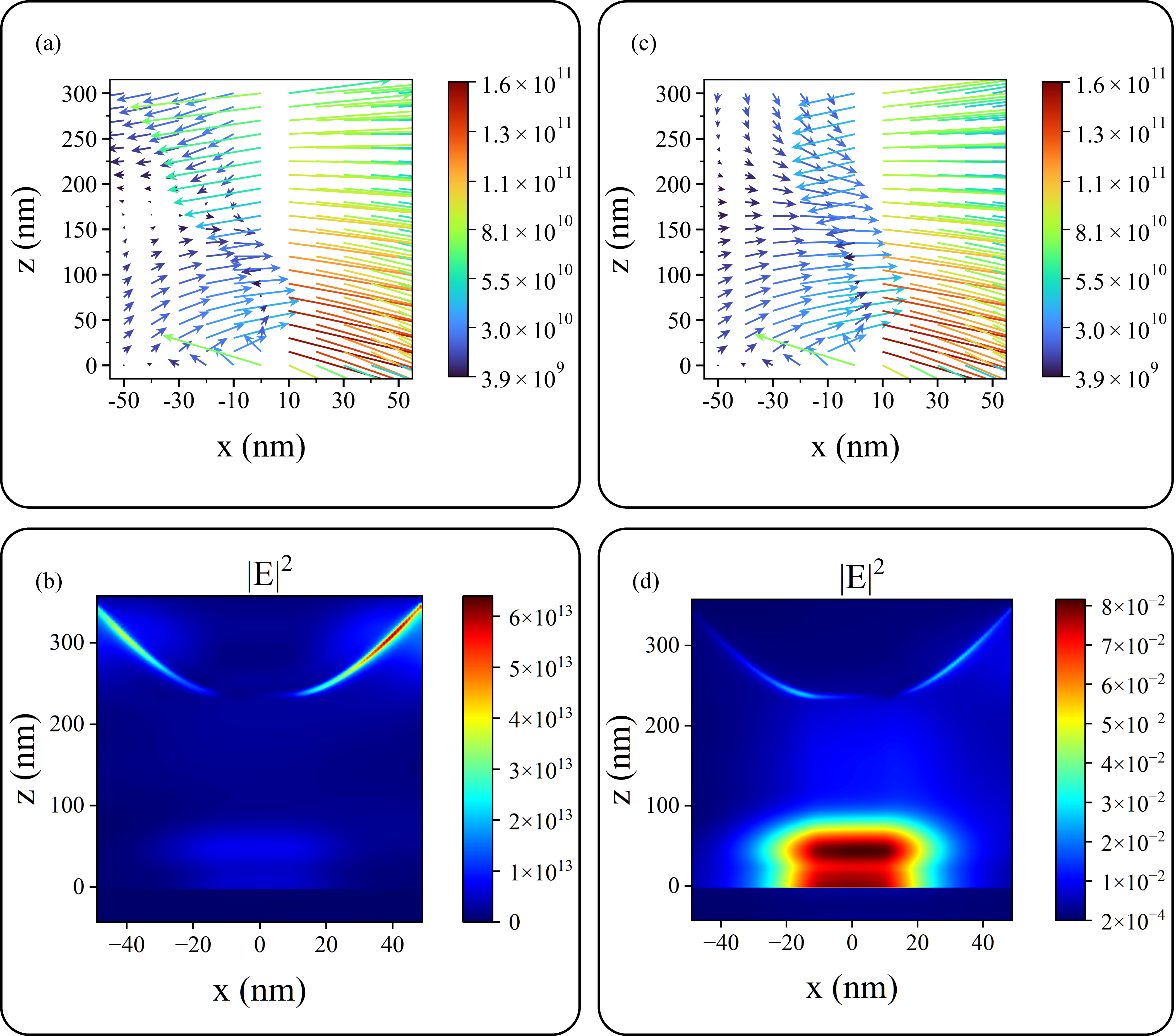}

    \caption{Computed optical properties of silicon semiconductor phase and metal phase. \textbf{(a)} A snapshot of the Poynting vector in the XZ-plane before metallization. \textbf{(b)} The field profile of the setup before metallization. \textbf{(c)} A snapshot of the poynting vector in the XZ-plane after metallization. \textbf{(d)}  The field profile of the setup after metallization.}
    \label{fig:4}
\end{figure}

The optical validations of silicon metallization have been illustrated in Fig.~\ref{fig:3} and Fig.~\ref{fig:4}. Fig.~\ref{fig:3} reveals highly negative real permittivity, high power absorption, large absorption coefficient, and a minuscule skin depth. Additionally, Fig.~\ref{fig:4} demonstrates circulating energy flux density (Poynting vectors) and reduced field activity ($|E|^2$) around the silicon tip. All features commonly associated with a typical metal \cite{ref17}. In our case, these features are extracted from a silicon undergoing metal phase transition. 

 Fig.~\ref{fig:3}a demonstrates a comparative analysis between the standard silicon semiconductor model and the silicon metallization that we performed. Standard silicon acts as a semiconductor. Hence, the real permittivity with respect to frequency is moderately positive, whereas the negative part, the dielectric loss factor, with respect to photon energy is negative \cite{ref23}. This is the expected behavior of a typical silicon semiconductor as shown by Fig.~\ref{fig:3}a. Conversely, metal follows the Drude model, where the real permittivity with respect to frequency is highly negative and the dielectric loss factor is also negative \cite{ref24}. Here, only the real permittivity sets metal and semiconductor apart. According to our result, silicon gives a metal-like real permittivity. Thus, it effectively behaves like a metal.

The comparative permittivity trace in Fig.~\ref{fig:3}(a) exhibits a clear qualitative signature of the Drude free-carrier response,
\begin{equation}
\varepsilon'(\omega) = \varepsilon_\infty - \frac{\omega_p^2}{\omega^2 + \gamma^2},
\label{eq:drude_eps}
\end{equation}
where $\varepsilon_\infty$ is the high-frequency background permittivity set by the bound (valence) electrons, $\gamma$ is the carrier scattering (damping) rate, and the plasma frequency $\omega_p = \sqrt{ne^2/\varepsilon_0 m^*}$ encodes the free-carrier density $n$ through the electron charge $e$, vacuum permittivity $\varepsilon_0$, and conductivity effective mass $m^*$ [24]. Using the FDTD-derived carrier density $n = 1.42 \times 10^{21}~\mathrm{cm^{-3}}$ (Eq.~B.8) and taking $m^{*} = 0.18\,m_0$ for the conductivity effective mass of silicon, the plasma frequency is calculated as
\begin{equation}
    \omega_p = \sqrt{\frac{n e^2}{\varepsilon_0 m^{*}}}
    = 5.01 \times 10^{15}~\mathrm{rad/s}
    \label{eq:omega_p_value}
\end{equation}
Combined with a carrier scattering rate $\gamma = 1.0 \times
10^{15}~\mathrm{rad/s}$ (Ref.~[17]), these parameters were used
within Eq.~(1) to generate the metallized-phase $\varepsilon'(\omega)$, $\varepsilon''(\omega)$ curves shown in Fig.~3(a). Physically, the $\omega_p^2/(\omega^2+\gamma^2)$ term is the free-carrier (Drude) contribution: it grows with carrier density $n$ and dominates $\varepsilon'$ whenever $\omega \ll \omega_p$, driving $\varepsilon'$ strongly negative, while it vanishes as $\omega \to \omega_p$, allowing $\varepsilon'$ to recover toward $\varepsilon_\infty$. For unexcited silicon, $\varepsilon'$ remains moderately positive and nearly frequency-independent, consistent with $n$ being low enough that the Drude term is negligible and $\varepsilon' \approx \varepsilon_\infty$ throughout. For the metallized case, $\varepsilon'$ instead drops sharply to large
negative values at low frequency and only recovers toward positive values as $\omega \to \omega_p$ -- exactly the qualitative signature Eq.~(\ref{eq:drude_eps}) predicts once $n$ crosses the Mott threshold and the Drude term becomes dominant. Crucially, this sign inversion and frequency-dependent recovery appear \emph{only} in the metallized curve and are entirely absent from the semiconductor curve, directly
confirming that the carrier population predicted analytically in Section~3.3 has physically driven the system into the Drude-dominated regime, rather than the negative permittivity being an isolated numerical feature.

 This notion is further solidified by Fig.~\ref{fig:3}b which depicts the absorption coefficient and field penetration depth. Particularly, field penetration depth
 gives us the definitive proof of metallization. Metals have exceedingly shallow field penetration depth. This is because metal skin depth is low that much of the incoming field is reflected or scattered at low angles \cite{ref24}. The aforementioned low skin depth is shown in Fig.~\ref{fig:3}d. On a different note, Fig.~\ref{fig:3}b also unveils the operating range of metallization. For roughly $800$~THz, the absorption coefficient of the metallic phase is well above the semiconductor phase. This indicates our metal phase lies somewhere between $1$~THz and $800$~THz. Another visual key to metallization is high conduction and high power absorption \cite{ref18}. Fig.~\ref{fig:3}c illustrates this perfectly. In Fig.~\ref{fig:3}c optical power is absorbed most efficiently near the apex point of the silicon cone, indicated by the $0.5$~nm curve which reaches an absorption of $220$~a.u. Such a high optical power absorption is typically seen in metals, thus raising the credibility of our work. Additionally, numerical work given in appendix~\ref{app:B} show an increase in the number of photons absorbed per pulse which also points toward a metal-like behavior \cite{ref17}.

In Fig.~\ref{fig:4}a, the Poynting vectors spread outward throughout the entire field of view. This is to be expected from a largely transparent medium such as a typical silicon semiconductor without any external optical stimulus. Power flows through and escapes the silicon rather than being reflected or trapped inside the material. This is due to the low absorption coefficient and the small extinction coefficient ubiquitously reported for intrinsic silicon \cite{ref23, ref24}.

On the other hand, in Fig.~\ref{fig:4}c, a subset of the vectors near the surface develops a circulating closed-loop trajectory instead of continuing the outward flow. This is a well-documented feature of a strongly reflective, high-loss interface; where incident and reflected (evanescent fields) components interfere to produce localized vortices in power flow \cite{ref25, ref26}. This behavior is commonly attributed to metals. Thus, this indicates a metal-like state in the silicon tip apex whilst keeping its semiconducting state intact in the farthest edges of the bulk material. This transient, spatially incomplete metallization is due to the presence of a dense, non-equilibrium electron-hole plasma  near the excited region rather than a fully metallic bulk state \cite{ref17}.

Figures~\ref{fig:4}(a) and \ref{fig:4}(c) reveal an unambiguous electromagnetic signature of the semiconductor-to-metal transition through the spatial topology of the energy flux
\begin{equation}
\mathbf{S} = \mathbf{E} \times \mathbf{H},
\label{eq:poynting}
\end{equation}
where $\mathbf{E}$ and $\mathbf{H}$ are the local electric and magnetic field vectors, so that $\mathbf{S}$ points in the direction of instantaneous power flow and its
magnitude gives the local energy flux density. A uniformly outward $\mathbf{S}$-field, as seen in Fig.~\ref{fig:4}(a), indicates power flowing away from the source and out
of the medium with minimal internal redirection -- the expected behavior for a largely transparent, low-absorption medium such as unexcited, phonon-limited
indirect-bandgap silicon. A closed-loop $\mathbf{S}$-field, as seen in Fig.~\ref{fig:4}(c), instead indicates that incident and reflected field components interfere to redirect power back into circulating trajectories rather than letting it
escape -- the expected topology at a strongly reflective, high-loss interface. The fact that this closed-loop behavior in Eq.~(\ref{eq:poynting}) appears only in a \emph{localized} region near the tip surface, while $\mathbf{S}$ farther from the
apex retains its original outward form, is the expected electromagnetic fingerprint of a localized metallic phase embedded in an otherwise semiconducting bulk, directly
corroborating the transient, spatially incomplete metallization picture developed in Sections~3.1 and~3.4, and confirming that the predicted carrier confinement volume $V_{\mathrm{eff}} \approx (\Delta r)^3$ (Eq.~(12)) -- where $\Delta r$ is the Au bump radius setting the size of the momentum-enhanced interaction region -- manifests as a correspondingly localized change in the simulated electromagnetic response.

In addition to the prior panels, Fig.~\ref{fig:4}b and Fig.~\ref{fig:4}d, demonstrate the contrasts between the field profiles of the normal semiconducting state and the transient metallic state of the transformed silicon tip. 

For Fig.~\ref{fig:4}b, $|E|^2$ is concentrated in a thin arc at and just beneath the surface, penetrating some depth into the silicon body. This is consistent with the existing literature for electric fields interacting with dielectric medium, where the field is only partially reflected and continues to propagate over a skin/absorption depth, set by the low absorption coefficient of unexcited silicon \cite{ref23, ref24}.

On the contrary, in Fig.~\ref{fig:4}d, $|E|^2$ collapses to a much lower magnitude at the silicon surface, indicating the field is predominantly reflected rather than transmitted. This is the perfect exemplar of a Drude-like behavior, where the real permittivity turns large and negative and the effective skin depth collapses \cite{ref17, ref24}. 

 The transition from Fig.~\ref{fig:4}(b) to Fig.~\ref{fig:4}(d) provides direct visual confirmation of the skin-depth collapse quantified independently in Fig.~\ref{fig:3}(d) via
\begin{equation}
\delta(\omega) = \frac{1}{\alpha(\omega)},
\label{eq:skin_depth}
\end{equation}
 where $\alpha(\omega)$ is the frequency-dependent absorption coefficient reported in Fig.~\ref{fig:3}(b), so that $\delta$ is the characteristic $1/e$ decay length of the field intensity into the material: a large $\alpha$ (strong absorption, as in the metallic phase) forces $\delta$ to be small, confining the field to a thin near-surface layer, while a small $\alpha$ (weak absorption, as in unexcited silicon) allows $\delta$ to be large, letting the field penetrate deep into the bulk. In the unexcited state (Fig.~\ref{fig:4}(b)), $|E|^2$ penetrates visibly deeper into the silicon body, consistent with the larger $\delta$ computed via Eq.~(\ref{eq:skin_depth}) for unexcited silicon over the same frequency range. In the metallized state (Fig.~\ref{fig:4}(d)), the field instead collapses into a shallow, near-surface region concentrated close to the tip base, consistent with the sharply reduced $\delta$ computed for the metallic phase in Fig.~\ref{fig:3}(d). These two figures are generated from entirely independent outputs of the same simulation --one a frequency-domain material property computed via Eq.~(\ref{eq:skin_depth}) (Fig.~\ref{fig:3}(d)), the other a real-space field-distribution snapshot (Fig.~\ref{fig:4}(d)) -- so their mutual agreement constitutes an internal consistency check that does not rely on any external reference data, reinforcing that the metallic-phase optical response is self-consistent across independently computed physical quantities.

 The frequency range over which the metallic-phase absorption coefficient $\alpha$ (Fig.~\ref{fig:3}(b)) exceeds the semiconductor-phase value coincides with the range over which the metallic-phase skin depth $\delta = 1/\alpha$ (Fig.~\ref{fig:3}(d), Eq.~(\ref{eq:skin_depth})) drops below the semiconductor-phase value, and both transitions onset near the same frequency at which the real permittivity $\varepsilon'$ (Fig.~\ref{fig:3}(a), Eq.~(\ref{eq:drude_eps})) crosses from strongly negative toward positive values. These three quantities are not independent physical observables but are analytically linked: $\varepsilon'$ sets the sign and magnitude of the Drude response, $\alpha$ follows from the imaginary part of the resulting complex refractive index $\tilde{n} = \sqrt{\varepsilon' + i\varepsilon''}$, and $\delta$ is simply the reciprocal of $\alpha$ through Eq.~(\ref{eq:skin_depth}). Their convergence onto the same frequency window, despite being plotted as three separately computed curves, demonstrates that the predicted metallic operating window is not an artifact isolated to a single derived quantity but emerges self-consistently across the full
 complex optical response of the simulated structure, providing a threefold internal validation of the metallization regime
central to the novelty of this work.

\subsection{Thermal Characteristics of Silicon Metallization}
\begin{figure}[H] 
    \centering
    \includegraphics[width=1\textwidth]{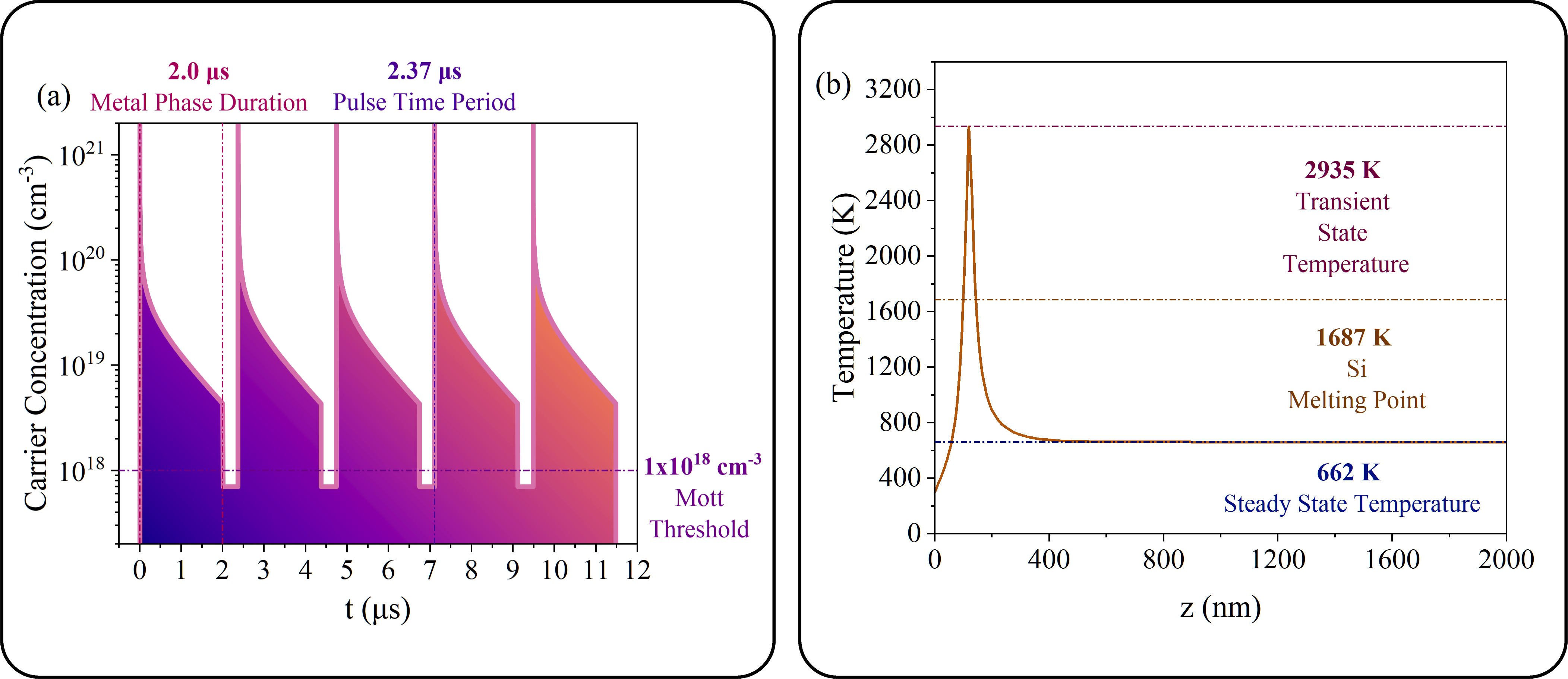}

    \caption{Computed thermal properties of Silicon semiconductor phase and metal phase.\textbf{(a)} Carrier concentration with respect to time showing metal phase duration and pulse period}\textbf{(b)} Localized temperature profile demonstrating transient and steady-state heat deposited along the z-direction as depicted in Fig.~\ref{fig:2}. 
    \label{fig:5}
\end{figure}

The nature of thermal states in silicon metallization and related carrier concentration that exceed Mott threshold is shown in Fig.~\ref{fig:5}a. The latter being a classic hallmark of silicon semiconductor undergoing metallization \cite{ref15}.

As mentioned before, the Mott threshold, or Mott criterion, refers to the critical point at which a semiconductor switches to a metal phase due to a high carrier concentration. This usually occurs when the valence and conduction bands overlap due to external excitations, meaning the number of charge carriers per unit volume across the first Brillouin zone increases drastically, effectively making a semiconductor to behave like a metal \cite{ref15,ref27}.In our work, the femtosecond pump photon energy exceeds silicon's indirect bandgap, so momentum-relaxed single-photon absorption (enabled by the Stage-1 confinement mechanism) injects carriers on a timescale faster than any recombination or thermal dissipation process, allowing the system to reach the metallic regime before it can relax  \cite{ref17,ref28}. Fig.~\ref{fig:5}a exemplifies this explanation further with the exact time duration of metal phase and carrier concentration. According to
Fig.~\ref{fig:5}a, the metal phase lasts for only 2.0~$\mu$s. The carrier population relaxes back below the Mott threshold over $\tau_{\mathrm{relax}} \approx 2.0~\mu$s (Fig.~5a); this timescale reflects carrier diffusion out of the confined $\Delta r \approx 2$~nm interaction volume rather than bulk recombination, since bulk Auger recombination alone at $n \sim 10^{21}$~cm$^{-3}$ would predict sub-nanosecond decay. While metal phase duration is
quite short, the pulse period is $2.37\,\mu\mathrm{s}$, giving a
duty cycle of $0.84$. At this $84\%$ duty cycle, the material spends most of each cycle in the metallic state, so across many consecutive cycles it exhibits pseudo-continuous metallic behavior, even though each individual metallic phase is transient. Furthermore, carrier concentration reaches well into $10^{21}~\mathrm{cm}^{-3}$, crossing the Mott threshold for silicon, $10^{18}~\mathrm{cm}^{-3}$ \cite{ref15}.

Fig.~\ref{fig:5}b shows the si-tip achieving a steady-state temperature of 662 K upon optical excitations. Although the transient peak (2935~K, Fig.~5(b)) exceeds silicon's
melting point, this value represents the peak spatial temperature at $t = 0^{+}$ immediately following pulse absorption; the corresponding temporal decay is estimated at $\sim 3$~ns, based on 1D thermal diffusion from the nanoscale-confined heated region ($\tau \approx L^2/4\alpha_{\mathrm{th}}$, with $\alpha_{\mathrm{th}} \approx
0.88~\mathrm{cm^2/s}$ for silicon), below the characteristic
melt-nucleation timescale ($\sim 10$~ns, Ref.~[32]), so rapid thermal relaxation prevents catastrophic failure, enabling long-term operation. Therefore, the pulsed beam does not irreparably damage the Si tip. This relates directly to the aforementioned phenomenon noticed in Fig.~\ref{fig:5}a where the electron-hole pair recombination and thermal dissipation process do not have enough time to notice the phase transition. Thus, mitigating any permanent damage to the material.

In relation to the points discussed earlier, as duty cycle reaches \(\sim 1\) pulsed laser behaves more like a CW laser. This leads to continuous heat deposition to the material and steady-state temperature exceed silicon’s natural melting point. Thus, leading to the Si tip melting. An apt way to circumvent this is to adjust the pulse period to trail behind the metal duration just by a small margin.

\subsection{Derived Mathematical model from FGR}

Fermi's golden rule (FGR) is a well known formula in quantum physics that offers the probability rate of bandgap transitions depending on the number of available density of states (DoS) and k-vectors \cite{ref20}.

For our purposes, we modify Fermi's golden (FGR) equation for interband transitions. Therefore, the transition rate from valence to conduction band is given by:

\begin{equation}
W_{vc}(k) \propto \frac{1}{V_e} |\mu_{cv}|^2 \int d^3q \exp\left[-\frac{q^2}{2\sigma_q^2}\right] \delta[\hbar\omega - E_{vc}(k,k+q)]
\end{equation}

where $V_e$ is the confined effective mode volume of photon, $\mu_{cv}$ is the transition dipole moment, $q$ is the photon wave vector, $\sigma_q$ is the standard deviation of photon momentum distribution, $E_{vc}$ is the energy difference between band states, and $\delta[\ldots]$ represents energy conservation through the delta function. The integral over $q$ allows transitions across momentum space, thereby enabling indirect transitions.

From the uncertainty principle, we have $\Delta k \cdot \Delta r \geq \pi$. For a photon confined in spatial size $\Delta r$, the momentum distribution is characterized by

\begin{equation}
\sigma_q \approx \frac{\pi}{\Delta r}
\end{equation}

As a numerical example, at $\Delta r = 2$ nm, we obtain $\sigma_q \approx \pi/(2\text{ nm}) \approx 1.57$ nm$^{-1}$. In comparison, for a free photon at wavelength $\lambda = 632.8$ nm, the momentum is $q_0 = 2\pi/\lambda \approx 0.01$ nm$^{-1}$. This yields an enhancement factor of $\sigma_q/q_0 \approx 157$.

    In addition, the effective absorption coefficient can be expressed as

\begin{equation}
\alpha_{\text{eff}} = \alpha_{\text{Si}} \times f(\Delta r, \lambda)
\label{eq:3}
\end{equation}

where the enhancement factor $f(\Delta r, \lambda)$ satisfies the limiting behaviors: $f \rightarrow 1$ when $\Delta r \gg \lambda$ (bulk Si, no enhancement) and $f \rightarrow \alpha_{\text{direct}}/\alpha_{\text{Si}}$ when $\Delta r \ll \lambda$ (direct bandgap behavior). The mathematical design philosophy behind Eq.~(\ref{eq:3}) and Eq.~(\ref{eq: 4}) is provided in appendix~\ref{app:A}.  A phenomenological model for this enhancement is:

\begin{equation}
f(\Delta r, \lambda) = 1 + \left(\frac{\alpha_{\text{direct}}}{\alpha_{\text{Si}}} - 1\right) \times \exp\left[-\left(\frac{\Delta r}{\Delta r_{\text{crit}}}\right)^2\right]
\label{eq: 4}
\end{equation}

where $\Delta r_{\text{crit}} \approx 0.8$ nm for $\lambda = 632.8$ nm \cite{ref20}, $\alpha_{\text{direct}} \approx 5 \times 10^7$ m$^{-1}$ (direct bandgap reference) and $\alpha_{\text{Si}} = 3.5 \times 10^5$ m$^{-1}$ (bulk Si at 632.8 nm).

The local field intensity at the picocavity is enhanced according to:

\begin{equation}
I_{\text{local}} = g^2 \cdot I_{\text{incident}}
\end{equation}

where $g$ is the field enhancement factor. From our simulation
results, we obtain $g \approx 10$ for a $\Delta r = 2$~nm Au bump.

As mentioned above, the setup involves dual lasers, namely, Gaussian laser and femtosecond pump pulse. The Gaussian CW laser acts as the momentum enabler. The femtosecond pulse acts as the carrier generator. 

For the Gaussian laser stage, the absorbed power per unit volume is given by $P_{\text{vol}} = \alpha_{\text{eff}} \cdot I_{\text{Gauss}}$, leading to the total absorbed power:

\begin{equation}
P_{\text{Gauss}} = \alpha_{\text{eff}} \cdot V_{\text{eff}} \cdot I_{\text{Gauss}}
\end{equation}

This creates an enhanced region where Si has $\alpha_{\text{eff}}$, effectively enabling `momentum-direct' transitions.

For the femtosecond pulse stage, the energy absorbed per pulse is:

\begin{equation}
E_{\text{pulse}} = \alpha_{\text{eff}} \cdot g^2 \cdot V_{\text{eff}} \cdot I_{\text{TFSF}} \cdot \tau_{\text{pulse}}
\end{equation}

where $\alpha_{\text{eff}}$ is the momentum-enhanced absorption, $g^2$ represents the intensity field enhancement, $I_{\text{TFSF}}$ is the peak intensity of the femtosecond pulse, $V_{\text{eff}} \approx (\Delta r)^3$ is the effective interaction volume, and $\tau_{\text{pulse}}$ is the pulse duration (1000 fs). The resulting carrier density generated is:

\begin{equation}
n_{\text{carrier}} = \frac{E_{\text{pulse}}}{V_{\text{eff}} \cdot \hbar\omega}
\end{equation}

Metallization occurs when $n_{\text{carrier}} > n_{\text{Mott}}$, leading to a transient metallic phase, where $n_{\text{Mott}} \approx 10^{18}$--$10^{20}$ cm$^{-3}$ represents the Mott transition density \cite{ref14,ref15,ref27}. This range is consistent with the Mott criterion $n^{1/3} a_B^* \approx 0.25$, where $a_B^* \approx 4.5$~nm is the exciton Bohr
radius in silicon, yielding $n_{\mathrm{Mott}} \approx 3.5 \times 10^{18}~\mathrm{cm^{-3}}$ \cite{ref27}, consistent with the lower bound of the adopted range. We have numerically verified this fundamental inequality in the appendix~\ref{app:B}.

The momentum enhancement scales as $\sigma_q \propto 1/\Delta r$ (from the uncertainty principle), while the enhancement factor follows the form given in Eq.~(4). The effective volume scales as $V_{\text{eff}} \propto (\Delta r)^3$ (picocavity volume), and the carrier generation follows:

\begin{equation}
n_{\text{carrier}} \propto \frac{\alpha_{\text{eff}} \cdot g^2 \cdot \tau_{\text{pulse}} \cdot I_{\text{TFSF}}}{(\Delta r)^3}
\end{equation}

Here, $V_{\mathrm{eff}}$ was approximated geometrically as
$(\Delta r)^3$ rather than computed via the rigorous mode-volume
integral $V_{\mathrm{eff}} = \int \varepsilon |E|^2 \, dV \, / \,\max\left(\varepsilon |E|^2\right)$; a full-field mode-volume calculation using the simulated $|E|^2$ distribution (Fig.~4) is left for future refinement.These competing effects reveal that smaller $\Delta r$ leads to larger $\alpha_{\text{eff}}$ (more absorption per volume), smaller $V_{\text{eff}}$ (less total volume), and larger $g^2$ (stronger field). The optimal $\Delta r$ for metallization is approximately 1--3 nm.

For a free space photon with $\lambda = 632.8$~nm, the momentum is $q_0 = 2\pi/\lambda \approx 0.01~\text{nm}^{-1}$, the volume is $V_0 = \lambda^3 \approx 2.53 \times 10^8~\text{nm}^3$, transitions are only vertical ($k' \approx k$), resulting in weak indirect absorption in Si. In contrast, for a confined photon with $\Delta r = 2$ nm, the momentum is $\sigma_q \approx \pi/\Delta r \approx 1.57$ nm$^{-1}$, the volume is $V_{\text{eff}} \approx (\Delta r)^3 \approx 8$ nm$^3$, transitions are diagonal ($\Delta k \approx \sigma_q$), resulting in strong direct-like absorption. The overall enhancement factor is approximately 10--100$\times$ depending on confinement. This carrier density represents a multi-pulse ensemble average; single-pulse shot noise (Appendix~B.2) implies individual pulses may transiently fall below $n_{\mathrm{Mott}}$, a limitation quantified further in Section~4.4.

At $t < 0$, the initial state consists of an Au nanostructure near a Si tip with $\Delta r = 2$ nm, where bulk Si exhibits properties characterized by $\alpha = \alpha_{\text{Si}}$ (indirect). At $t = 0^+$, the Gaussian laser is turned on, providing continuous confinement at the picocavity with photon momentum $\sigma_q \approx \pi/\Delta r$. The Si locally becomes characterized by $\alpha \rightarrow \alpha_{\text{eff}} \approx 10\alpha_{\text{Si}}$, effectively creating a `direct bandgap' region with baseline heating of $\Delta T_{\text{base}} \approx$ hundreds of K.

At $t = t_{\text{pulse}}$, a femtosecond pulse (1000~fs) arrives, delivering high peak intensity on the `direct' Si region. This results in massive carrier generation with $n \approx 1.42\times10^{21}$~cm$^{-3}$, where $n > n_{\mathrm{Mott}}$ leads to transient metallization lasting approximately femtoseconds, with pulse heating of $\Delta T_{\mathrm{pulse}} \approx$ a few K per event (owing to the short single-pulse energy deposition). Note that this femtosecond timescale describes the carrier-\emph{injection} process within a single pulse; the metallic phase itself, sustained by the injected carrier population, persists over the much longer $\tau_{\mathrm{relax}} \approx 2.0~\mu$s decay time
discussed in Section~3.2 and shown in Fig.~\ref{fig:5}a.

At $t = t_{\mathrm{pulse}} + \tau_{\mathrm{relax}}$, the carrier population relaxes back below the Mott threshold over $\tau_{\mathrm{relax}} \approx 2.0~\mu$s (Fig.~5a), reflecting carrier diffusion out of the confined $\Delta r \approx 2$~nm interaction volume rather than bulk recombination, since bulk Auger recombination alone at $n \sim 10^{21}$~cm$^{-3}$ would predict sub-nanosecond decay, and Si returns to the 'direct' (not bulk) state while the Gaussian laser still maintains momentum enhancement.At $t =1/f_{\text{rep}} \approx  2.37~\mu\text{s}$ (next pulse), the process repeats at $\sim 422$~kHz with stable cycling between metallization and direct Si, while the steady-state temperature remains well below melting (Fig.~\ref{fig:5}b). The entire set of equations governing such phenomena can be found in the supplementary material.

\subsection{Analysis at Atomic Scales}
Considering all the prior discussed results, at the atomic level, the femtosecond pump photon energy ($\hbar\omega \approx 1.96$~eV, $\lambda = 632.8$~nm) exceeds silicon's indirect bandgap ($E_g \approx 1.12$~eV), so energy conservation does not forbid single-photon absorption, the obstacle is momentum conservation. The photon-momentum confinement established in Stage~1 bridges this momentum mismatch ($\sigma_q \approx \pi/\Delta r$), enabling phonon-free absorption within the confined region. The femtosecond pulse (Stage~2) then injects carriers through this channel fast enough to outpace recombination and lattice thermalization, consistent with reported wavelength- and intensity-dependent carrier-injection dynamics in silicon \cite{ref28}. As the pulse duration is considerably shorter than the electron-phonon scattering time ($~100$ fs–1 ps) and far shorter than carrier recombination lifetimes, electrons are pumped into the conduction band, and holes into the valence band, much faster than the lattice can respond thermally or the plasma can recombine, producing a highly non-equilibrium, dense electron-hole plasma confined to the excited volume \cite{ref17}. At this stage, the silicon crystal still retains its diamond-cubic lattice, but its electronic subsystem has been driven into a state with free-carrier densities approaching $10^{21}$~cm$^{-3}$,
several orders of magnitude above the thermal-equilibrium value.

As carrier density climbs, the Coulomb interaction between electrons and holes, which normally binds them into excitons, becomes progressively screened by the surrounding delocalized free-carriers. Once the carrier concentration surpasses the critical Mott density, the exciton binding energy collapses to zero. This means, bound electron-hole pairs can no longer exist as discrete quasiparticles, and the system undergoes an exciton Mott transition into an unbound, degenerate electron-hole plasma \cite{ref15}. This threshold is a specific case of the more general Mott criterion, where the criterion for the insulator-to-metal crossover is set by the interplay between the reduced Bohr radius of the bound state and the mean interparticle spacing of the free-carrier gas. Once carriers are packed closer together than this critical spacing, delocalized (metallic) behavior triumphs over any localized/bound electronic configuration \cite{ref27}. Additionally, the sheer density of conduction band electrons and valence band holes renormalizes the band structure itself. This means, exchange-correlation effects among the dense carriers shrink the fundamental gap (bandgap renormalization) and pushes the silicon's electronic structure toward a kind of continuous, partially-filled band~\cite{ref10}.

Once the electron-hole plasma is dense and unbound, its optical response is no longer dictated by interband (semiconductor) transitions but by the free, quasi-classical oscillation of carriers in the applied field. Therefore, effectively, the transformed silicon obeys the same physics that describes conduction electrons in a metal, captured by the Drude model \cite{ref24}. A step-by-step atomic process sequence can be found in the supplementary material.

\section{Experimental Setup Feasibility and Tolerance Analysis}
\label{possible_approaches}
 Even though the setup aforementioned is theoretical, it can be realized on an experimental platform which employs a gap-mode plasmonic architecture consisting of a 500~$\mu$m SiO$_2$ substrate (chosen for practicality), 50~nm  $\text{Au}$ film, 2~nm  $\text{Au}$ hemispherical bump, and \SI{2}{\micro\meter} crystalline  $\text{Si}$ tip positioned under piezoelectric feedback control. This nanoparticle-on-mirror (NPoM) configuration confines the electromagnetic hot spot within the tip bump junction, with gap distances controllable to sub-nanometer precision range \cite{ref25}. Dual-beam excitation, a Gaussian beam for steady-state driving and a synchronized femtosecond pump pulse, delivered by the Ti:Sapphire CPA system described below, illuminates the junction at an oblique angle, directly mirroring the simulation conditions discussed earlier \cite{ref30}. As such, this configuration enables simultaneous gap-mode resonance excitation and background free collection of the near-field scattered signal, essential for isolating nanoscale Raman and scattering responses from far-field contributions.

Although, a physical experimental setup has not been built for this work, the latter sub-sections provide deeper insights into such a setup—if ever practically built. A detailed discussion of candidate characterization techniques, including gap-mode TERS for carrier-density-sensitive phonon lineshape analysis, variable-angle spectroscopic ellipsometry for direct  extraction, femtosecond pump-probe reflectometry for transient carrier dynamics, and Hall-effect measurements for direct carrier quantification, together with the specific measurable signatures each technique would produce if the predicted transient metallization occurs, is provided in the Supplementary Material (Section 3). Thus, we explore its practical setup feasibility and with a brief tolerance analysis. The real-world analytical procedures are detailed in the  supplementary material. 

\subsection{Basic Layer Fabrication}
For the \SI{500}{\micro\meter} $\text{SiO}_2$ glass substrate dry thermal oxidation is used at \SI{1323}{\kelvin}, yielding a stoichiometric oxide with sub-nanometer roughness. This is further verified by AFM and ellipsometry. Afterward, the 50~nm $\text{Au}$ film is deposited via electron-beam evaporation $< 5 \times 10^{-7}~\text{Torr}$ with a 3 nm $\text{Ti}$ adhesion layer, maintaining an RMS roughness below 0.5 nm to prevent parasitic plasmonic hotspots which would obscure the targeted junction response \cite{ref27}. Additionally, film thickness is confirmed by cross-sectional TEM and XRR. Lastly, the \SI{2}{\micro\meter} $\text{Si}$ tip is acquired from commercial n-type AFM cantilevers, with SEM-verified apex radii $< 10~\text{nm}$. Strategic n-doping positions the equilibrium Fermi level near the conduction band, reducing the carrier injection threshold for the Mott transition and amplifying the optical sensitivity to density variations. 

\subsection{Complex Fabrication of the 2~nm \texorpdfstring{$\text{Au}$}{Au} Bump}
For this experimentation, the 2 nm $\text{Au}$ bump poses the primary fabrication problem. This is because conventional lithography cannot precisely give such a fine resolution. Three complementary routes are considered. Firstly, thermal dewetting of a 0.3–0.5~nm $\text{Au}$ film at \SIrange{423}{473}{\kelvin} drives spontaneous island nucleation, with bump height tuned via deposited mass and annealing duration. Secondly, Atomic layer deposition (ALD) followed by electroless plating yields isolated particles with sub-nanometer gap control. This is fully compatible with the $\text{Au}$/$\text{SiO}_2$ stack \cite{ref26, ref29}; related ALD, EBL, and FIB methods have reliably achieved sub-10~nm metallic gaps \cite{ref31,ref32}. Thirdly, Controlled electrochemical deposition from dilute $\text{HAuCl}_4$ enables precise bump nucleation, with height verified through in-situ STM. Regardless of the chosen route, final structures are characterized by cross-sectional HRTEM and dark-field scattering spectroscopy, which can non-destructively confirm the expected NPoM gap-mode resonance \cite{ref33}.

Additionally, photon momentum confinement to nano-scaled structures for bulk silicon has been proposed, designed, and physically tested under lab conditions, but for different outcomes \cite{ref34,ref35}. Nevertheless, such setup can be taken as benchmarks when designing the physical experiment for the proposed setup. 

\subsection{Optical Excitation Sources \& Laser Systems}
For source excitations, three optical excitation modalities are integrated into the setup. A Gaussian beam (CW or ns-pulsed) is fiber delivered and focused through a long working distance objective (NA 0.7–0.8) for steady-state Raman and scattering measurements. A synchronized femtosecond pump pulse, derived from the same Ti:Sapphire source via a beamsplitter and delay arm, illuminates at an oblique angle for background isolated NF probing in backscattering geometry. The principle carrier-injection driver is a Ti:Sapphire chirped-pulse amplifier (632.8~nm, 1000~fs, 422~kHz). The well known Three-Temperature Model (3TM) is expected, based on prior literature, to predict a reversible metallization window for near-IR femtosecond excitation, bounded by Mott onset and the silicon damage threshold ($0.65$~J/cm$^2$) \cite{ref28}; a dedicated 3TM analysis for the present geometry remains to be performed.

\subsection{Tolerance Analysis}
The tolerance boundaries observed across parameters follow directly from the underlying carrier dynamics. At the upper intensity limit of the Gaussian conditioning beam, above $I_{\mathrm{Gauss}} \approx 5\times10^{10}$~Wm$^{-2}$ (Table 1), the rate of hot-carrier generation exceeds what the lattice and carrier system can relax through normal electron-phonon and recombination channels, driving an avalanche-like build-up of carrier density that pushes the system into an unrealistic, over-driven absorption regime rather than a resolvable transient metallic state \cite{ref12}. Consequently, at the lower bound ($I_{\mathrm{Gauss}} \approx 2.4\times10^{1}$~Wm$^{-2}$), generation never reaches the critical density needed to cross the Mott threshold, so the medium remains semiconducting throughout \cite{ref15}. 

Additionally, the $84\%$ duty cycle appears optimal. This is because, across many consecutive pulse cycles, the material spends the majority of each cycle in the metallic state, so even though each individual metallization event is transient and decays before the next pulse, the accumulated effect across cycles produces an apparent pseudo-continuous metallic response rather than a series of isolated spikes.

Lastly, the $2$--$3$~nm gold bump radius window reflects the
near-field enhancement provided by the nanostructure itself. At
this scale, the bump concentrates the incident field enough to
locally boost carrier generation and drive metallization. Below
$2$~nm, the concentrated field causes thermal and structural damage
to the bump and the silicon tip, whereas above $3$~nm the
enhancement falls off enough that the bump no longer meaningfully
alters the excitation relative to the unstructured case
\cite{ref20}

\section{Conclusion}
\label{conclusion}
The indirect bandgap nature of semiconductor materials like silicon has been a major obstacle since the inception of the electronics and photonics research base. Prior research suggests bandgap manipulation can be achieved through strain, field-induced electronic nonlinearity, and magnetic-field-driven electronic transitions. Building on the established photon-momentum confinement mechanism stated by Kharintsev et al.~\cite{ref20} and Noskov et al.~\cite{ref34,ref35} but deviating from their respective conclusions, our work's central contribution is the subsequent carrier-density-driven stage: sequentially modifying the bandgap beyond simple transitions into full-on, reversible metallization whilst preventing excess heating. This work demonstrates metallization in silicon with the aid of femtosecond laser. Firstly, a CW Gaussian laser is used to transition into a direct bandgap zone. Secondly, a pulsed femtosecond laser is used to drive the system through the forbidden region, giving us a transient metal-like behavior. Lastly, the pulse period is modulated to shadow the metal phase duration to obtain a near-perfect metallic state.

The method of metallization proposed in this paper is relatively nascent. That being said, this method may offer a rich tapestry of designs and industrial applications for the near future; much of  industry is based on silicon, and silicon is at the heart of this paper. Photon momentum induced direct bandgap formation in silicon combined with electric field driven metallization could enable a versatile platform for next-generation optoelectronic and electronic systems. By overcoming the inherent indirect bandgap of silicon active photonic components including lasers, modulators and amplifiers may be monolithically integrated into conventional CMOS technology. Wavelength tunable, voltage responsive silicon based light emitters such as LEDs and laser diodes can be made possible by dynamic modulation of the electronic band structure by optical stimulation and external bias. In addition to photonics, this dual stimulus method may enable multi-stimulus computing architectures, hybrid electro-optical logic and electrically and optically adjustable circuit elements without the need for mechanical switching or permanent doping. When combined, field-light-driven phase modulation can potentially create a new paradigm for integrated photonic and unconventional computing systems, transforming silicon from a passive electronic material into an actively reconfigurable platform.

\textbf{Supplementary Material}\\[0.3cm]
See the supplementary material for the derivation of the 
interpolation model, associated calculations, and numerical 
verification of the photo-generated carrier density exceeding 
the Mott transition threshold—along with a discussion on possible experimental verification and characterization setup.

\textbf{Acknowledgments}\\[0.3cm]
M.\,R.\,C.\ Mahdy acknowledges the support of the NSU (North South University) internal grants: CTRG grant 2024--25 and CTRG grant 2025--26 (approved by the members of BOT, NSU, Bangladesh).

\textbf{Disclosures}\\[0.3cm]
The authors declare no conflicts of interest.

\textbf{Data Availability}\\[0.3cm]
Data supporting the findings of this study are available in the article and its supplementary material.

\newcounter{appendixcounter}
\renewcommand{\theappendixcounter}{\Alph{appendixcounter}}

\refstepcounter{appendixcounter}
\section*{Appendix \theappendixcounter. Derivation of Interpolation Model}
\label{app:A}
\addcontentsline{toc}{section}{Appendix \theappendixcounter. Derivation of Interpolation Model}
\renewcommand{\thesubsection}{\theappendixcounter.\arabic{subsection}}
\renewcommand{\theequation}{\theappendixcounter.\arabic{equation}}
\setcounter{subsection}{0}
\setcounter{equation}{0}

Consider $\alpha_{\text{Si}}$ as standard silicon absorption coefficient. This represents the natural absorption coefficient of bulk crystalline silicon for indirect optical transitions (phonon-assisted). Silicon exhibits a small absorption coefficient because it has an indirect bandgap at 1.1 eV. Transitions require electron + photon + phonon (a 3-body process), which makes absorption approximately 100$\times$ weaker than direct bandgap materials \cite{ref23}.

At wavelength $\lambda = 632.8$ nm (photon energy $E = 1.96$ eV), the numerical value is
\begin{equation}
\alpha_{\text{Si}} = 3.5 \times 10^5 \text{ m}^{-1} = 3500 \text{ cm}^{-1}
\end{equation}

\textbf{This means;}

\begin{equation}
\text{Penetration Depth} = \frac{1}{\alpha_{\text{Si}}} = 2.86\,\mu\text{m}.
\end{equation}
Consider $\alpha_{\text{direct}}$ as direct bandgap reference absorption. This is the absorption coefficient of a direct bandgap semiconductor, used as a reference for the maximum possible enhancement \cite{ref20, ref28}. At wavelength $\lambda = 632.8$ nm (photon energy $E = 1.96$ eV), the numerical value is
\begin{equation}
\alpha_{\text{direct}} = 5 \times 10^7 \text{ m}^{-1} = 5 \times 10^5 \text{ cm}^{-1}
\end{equation}
\textbf{The comparison shows} 
\begin{equation}
    \alpha_{\text{target}}=\alpha_{\text{direct}}/\alpha_{\text{Si}} = 142.9
\end{equation}
\textbf{Here, $\boldsymbol{\alpha}_{\text{target}}$ represents the target enhancement from the photon momentum.}
\vspace{6pt}

Consider $\alpha_{\text{eff}}$ as effective absorption coefficient. This is the modified absorption coefficient of silicon when photons are confined to nanometer scales. The definition in our model is stated as
\begin{equation}
\alpha_{\text{eff}}(\Delta r) = \alpha_{\text{Si}} \times f(\Delta r, \lambda)
\end{equation}

In the Photon momentum based literature, it is stated that absorption enhancement can be achieved  through a different route, namely, transforming $P = \alpha_{\text{direct}} \times V \times I$ through the effective change of the absorption coefficient $\alpha_{\text{direct}}$.Moreover, prior research demonstrates that Si behaves as if it has $\alpha_{\text{direct}}$ when confined \cite{ref20}. However, to the best of our knowledge, no study has explicitly applied the $\alpha_{\text{eff}}(\Delta r)$ formula seen in our model.
\vspace{6pt}

Our experiments indicate that 1~nm structures exhibit massive
heating leading to tip melting, 2~nm structures show strong
heating, 3~nm structures show moderate heating, and 5~nm
structures show minimal heating. This size-dependent near-field
response is consistent with our earlier report on Au-based
nanostructures, where the interaction volume and field
confinement around comparably sized gold features were
similarly found to govern the strength of the resulting optical
response for Rayleigh-scale objects \cite{ref36}. This shows a clear inverse relationship, where smaller $\Delta r$ leads to more absorption \cite{ref20,ref37,ref38}.
\vspace{6pt}

\textbf{We model this as}
\begin{equation}
\alpha_{\text{eff}} = \alpha_{\text{Si}} \times f(\Delta r, \lambda)
\end{equation}

\textbf{Here, \textbf{$f$} captures the momentum-enabled enhancement. }
\vspace{6pt}

Consider $g$ as Field Enhancement Factor. This represents the enhancement of the electric field amplitude from plasmonic resonance, where $E_{\text{local}} = g \cdot E_{\text{incident}}$ and $I_{\text{local}} = g^2 \cdot I_{\text{incident}}$ (intensity enhancement). Such plasmonic-resonance-driven field enhancement over gold nanostructures has
separately been exploited to generate distinct, particle-type-dependent optical forces for single-particle sorting \cite{ref36}, and, in a metasurface-integrated form, to sustain broadband optical pulling across heterogeneous nanoparticle clusters \cite{ref39}.

\vspace{6pt}


\refstepcounter{appendixcounter}
\section*{Appendix \theappendixcounter. Calculations Based on Interpolation Model}
\label{app:B}
\addcontentsline{toc}{section}{Appendix \theappendixcounter. Calculations Based on Interpolation Model}
\renewcommand{\thesubsection}{\theappendixcounter.\arabic{subsection}}
\renewcommand{\theequation}{\theappendixcounter.\arabic{equation}}
\setcounter{subsection}{0}
\setcounter{equation}{0}

The material of interest, silicon, is excited in two stages. Stage one involves the application of the Gaussian CW laser, and stage two involves the application of a pulsed femtosecond beam. The input parameters are as follows:
\begin{table}[!htbp]
    \centering
    \begin{tabular}{|p{6cm}|c|}\hline
         \multicolumn{2}{|c|}{\textbf{Input Parameters}}\\\hline
         Photon energy, $E = \hbar\omega$ 
& 
    $1.9593\,\text{eV}$ \\\hline 
 Gaussian intensity, $I_{\text{Gauss}}$ 
&$5 \times 10^{10}\,\text{Wm}^{-2}$ \\\hline
Pulsed beam peak intensity, $I_{\text{TFSF}}$
&$1000\times10^{10}\,\text{Wm}^{-2}$ \\\hline
 Pulse width, $\tau$ 
&$1000\,\text{fs}$ \\\hline
 Pulse repetition rate, $f_{\text{rep}}$ 
&$\sim 422\,\text{kHz}$ \\\hline
 Au bump size, $\Delta r$ 
&$2\,\text{nm}$ \\ \hline\end{tabular}
    \caption{List of Input Parameters.}
    \label{tab:parameters}
\end{table}
\subsection{Application of the Gaussian Beam}
Based on the input parameters; momentum enhancement factor is $f(2 \text{ nm}, 632.8 \text{ nm}) = 1.274$. Therefore, the effective absorption coefficient becomes
\begin{equation}
\alpha_{\text{eff}} = \alpha_{\text{Si}} \times f(\Delta r, \lambda) = 3.5 \times 10^5 \times 1.274 = 4.458 \times 10^5 \text{ m}^{-1}
= 4\,458\,\text{cm}^{-1}
\end{equation}
\textbf{The effective mode volume is given as}

\begin{equation}
V_{\text{eff}} \approx (\Delta r)^3 = (2 \text{ nm})^3 = 8.0 nm^3 = 8.0 \times 10^{-27} m^3 
\end{equation}

\textbf{Therefore, the Gaussian CW absorption is shown as}

\begin{equation}
P_{\text{Gauss}} = \alpha_{\text{eff}} \times I_{\text{Gauss}} \times V_{\text{eff}} = 178.3 \text{ pW}
\end{equation}

\textbf{This continuous absorption creates the `direct' Si region.}

\subsection{Application of the Femtosecond Pulsed Beam}

Through our analytical work , we get $g = 10$, giving $g^2 = 100$, for a 2 nm Au bump. Therefore, the local intensity becomes:
\begin{equation}
I_{\text{local}} = g^2 \times I_{\text{TFSF}} = 100 \times 1000 \times 10^{10} = 1.0 \times 10^{15} \text{ W/m}^2
\end{equation}

\textbf{The energy per pulse (application of both beams considered) given by the equation}
\begin{equation}
E_{\text{pulse}} = \alpha_{\text{eff}} \times I_{\text{local}} \times V_{\text{eff}} \times \tau_{\text{pulse}} = \alpha_{\text{eff}} \times (g^2 \times I_{\text{TFSF}}) \times V_{\text{eff}} \times \tau_{\text{pulse}}
\end{equation}

\textbf{This yields as}
\begin{equation}
E_{\text{pulse}} = 3.567 aJ = 3.567 \times 10^{-18} J
\end{equation}

\textbf{Therefore, the number of photons absorbed per pulse yields}
\begin{equation}
N_{\text{photons}} = \frac{E_{\text{pulse}}}{\hbar\omega} = \frac{3.567 \text{ aJ}}{3.139 \times 10^{-19} \text{ J}} = 11.36 \text{ photons}
\end{equation}

\textbf{The generated carrier density is given by}
\begin{equation}
n_{\text{carrier}} = \frac{N_{\text{photons}}}{V_{\text{eff}}} = 1.420 \times 10^{27} \text{ m}^{-3} = 1.420 \times 10^{21} \text{ cm}^{-3}
\end{equation}

For the metallization check, the Mott transition threshold is $10^{18}$--$10^{20}$ cm$^{-3}$ \cite{ref14,ref15,ref27}. The generated carrier density is $1.420 \times 10^{21}$ cm$^{-3}$. Therefore, metallization occurs since $n > n_{\text{Mott}}$. The enhancement breakdown shows; momentum enhancement
$f(2~\mathrm{nm}) = 1.274$, field enhancement $g^2 = 100$, and total absorption boost $f \times g^2 = 127.4$. This matches with our observation of metallization at 2 nm. The analytical estimate in Eq.~(B.8) and the FDTD-derived absorbed power in Eq.~(B.3) share common input parameters ($\alpha_{\mathrm{eff}}$, $V_{\mathrm{eff}}$, $I$) rather than constituting independent measurements; the agreement
therefore reflects internal consistency of the model rather than
independent cross-validation. Independent validation of
$n_{\mathrm{carrier}}$ --- e.g., via a rate-equation model fit
directly to the FDTD-extracted absorbed power spectrum --- is
identified as necessary future work (Section~4.4). The mean number of absorbed photons per pulse, $N_{\mathrm{photons}} \approx 11.4$ (Eq.~B.7), is small enough that single-pulse absorption is subject to appreciable Poissonian fluctuation, with a relative standard deviation of $\sigma_N/N_{\mathrm{photons}} \approx 1/\sqrt{N_{\mathrm{photons}}} \approx 30\%$. This shot-noise level implies that any individual pulse may generate a carrier density that deviates from the mean value reported in Eq.~B.8 by a comparable fraction. However, because the system is driven at $f_{\mathrm{rep}} \approx 422$~kHz with an 84\% duty cycle (Section~3.2), the reported metallic-phase duration and steady-state thermal behavior (Fig.~5) reflect the time-averaged response over many consecutive pulses rather than a single-shot event. The mean carrier density $n_{\mathrm{carrier}}$ used to assess the Mott criterion should therefore be interpreted as a multi-pulse ensemble average; the single-pulse fluctuation does not affect the validity of the steady-state and duty-cycle conclusions, although, it does imply that the metallization threshold may be transiently undercrossed on individual pulses within the train.

\clearpage
\phantomsection


\addcontentsline{toc}{section}{References}
\begin{thebibliography}{99}

\bibitem{ref1}
Low J J, Kreider M L, Pulsifer D P, Jones A S and Gilani T H BAND GAP ENERGY IN SILICON 7

\bibitem{ref2}
Grundmann M 2016 The Physics of Semiconductors: An Introduction Including Nanophysics and Applications (Cham Heidelberg New York Dordrecht London: Springer)

\bibitem{ref3}
Precker J W and Da Silva M A 2002 Experimental estimation of the band gap in silicon and germanium from the temperature–voltage curve of diode thermometers American Journal of Physics 70 1150–3

\bibitem{ref4}
Sun G 2011 Intersubband approach to silicon based lasers—circumventing the indirect bandgap limitation Adv. Opt. Photon. 3 53

\bibitem{ref5}
Wolfe T ELECTRONIC AND OPTICAL PROPERTIES OF FIRST-ROW TRANSITION METALS IN 4H-SIC FOR PHOTOCONDUCTIVE SWITCHING

\bibitem{ref6}
Li S, Chou J-P, Zhang H, Lu Y and Hu A 2019 A study of strain-induced indirect-direct bandgap transition for silicon nanowire applications Journal of Applied Physics 125 082520

\bibitem{ref7}
Patel S, Dey U, Adhikari N P and Taraphder A 2022 Electric field and strain-induced band-gap engineering and manipulation of the Rashba spin splitting in Janus van der Waals heterostructures Phys. Rev. B 106 035125

\bibitem{ref8}
Higuchi K, Hamal D B and Higuchi M 2022 Second-order phase transition of silicon from a band insulator to metal induced by strong magnetic fields New J. Phys. 24 103028

\bibitem{ref9}
Cao T, Wang D, Geng D-S, Liu L-M and Zhao J 2016 A strain or electric field induced direct bandgap in ultrathin silicon film and its application in photovoltaics or photocatalysis Phys. Chem. Chem. Phys. 18 7156–62

\bibitem{ref10}
Hennig R G, Wadehra A, Driver K P, Parker W D, Umrigar C J and Wilkins J W 2010 Phase transformation in Si from semiconducting diamond to metallic $\beta$-Sn phase in QMC and DFT under hydrostatic and anisotropic stress \textit{Phys. Rev. B} \textbf{82} 014101

\bibitem{ref11}
Lin L, Li Z, Feng J and Zhang Z 2013 Indirect to direct band gap transition in ultra-thin silicon films Phys. Chem. Chem. Phys. 15 6063

\bibitem{ref12}
Turkulets Y and Shalish I 2018 Franz-Keldysh effect in semiconductor built-in fields: Doping concentration and space charge region characterization Journal of Applied Physics 124 075102

\bibitem{ref13}
Fukuoka S, Oka T, Ihara Y, Kawamoto A, Imajo S and Kindo K 2024 Magnetic field induced insulator-to-metal Mott transition in $\lambda$ -type organic conductors Phys. Rev. B 109 195142

\bibitem{ref14}
Corkish R, Chan D S-P and Green M A 1996 Excitons in silicon diodes and solar cells: A three-particle theory Journal of Applied Physics 79 195–203

\bibitem{ref15}
Suzuki T and Shimano R 2012 Exciton Mott Transition in Si Revealed by Terahertz Spectroscopy Phys. Rev. Lett. 109 046402

\bibitem{ref16}
Palik E D and Ghosh G 1998 Handbook of optical constants of solids (San Diego: Academic Press)

\bibitem{ref17}
Sokolowski-Tinten K and Von Der Linde D 2000 Generation of dense electron-hole plasmas in silicon Phys. Rev. B 61 2643–50

\bibitem{ref18}
Tateda M, Iida Y and Miyaji G 2023 Enhancement of plasmonic coupling on Si metallized with intense femtosecond laser pulses Sci Rep 13 18414

\bibitem{ref19}
Liu Y, Ding Y, Xie J, Chen M, Yang L, Lv X and Yuan J 2022 Research on Monocrystalline Silicon Micro-Nano Structures Irradiated by Femtosecond Laser Materials 15 4897

\bibitem{ref20}
Kharintsev S S, Noskov A I, Battalova E I, Katrivas L, Kotlyar A B, Merham J G, Potma E O, Apkarian V A and Fishman D A 2024 Photon Momentum Enabled Light Absorption in Silicon \textit{ACS Nano} \textbf{18} 26532–40

\bibitem{ref21}
Yamaguchi M and Nobusada K 2016 Indirect interband transition induced by optical near fields with large wave numbers Phys. \textit{Rev. B} \textbf{93} 195111

\bibitem{ref22}
Noda M, Iida K, Yamaguchi M, Yatsui T and Nobusada K 2019 Direct Wave-Vector Excitation in an Indirect-Band-Gap Semiconductor of Silicon with an Optical Near-field Phys. \textit{Rev. Applied} \textbf{11} 044053

\bibitem{ref23}
Green M A and Keevers M J 1995 Optical properties of intrinsic silicon at 300 K \textit{Progress in Photovoltaics} \textbf{3} 189–92

\bibitem{ref24}
Hui R and O’Sullivan M 2009 Fundamentals of Optical Devices Fiber Optic Measurement Techniques (Elsevier) pp 1–128

\bibitem{ref25}
Sigle D O, Mertens J, Herrmann L O, Bowman R W, Ithurria S, Dubertret B, Shi Y, Yang H Y, Tserkezis C, Aizpurua J and Baumberg J J 2015 Monitoring Morphological Changes in 2D Monolayer Semiconductors Using Atom-Thick Plasmonic Nanocavities ACS Nano 9 825–30

\bibitem{ref26}
Jeong H-H, Adams M C, Günther J-P, Alarcón-Correa M, Kim I, Choi E, Miksch C, Mark A F, Mark A G and Fischer P 2019 Arrays of Plasmonic Nanoparticle Dimers with Defined Nanogap Spacers ACS Nano 13 11453–9

\bibitem{ref27}
Pergament A, Stefanovich G and Markova N The Mott criterion: So simple and yet so complex

\bibitem{ref28}
Venkat P and Otobe T 2022 Wavelength dependence of laser-induced excitation dynamics
in silicon Appl. Phys. A 128 810

\bibitem{ref29}
Mubeen S, Zhang S, Kim N, Lee S, Krämer S, Xu H and Moskovits M 2012 Plasmonic Properties of Gold Nanoparticles Separated from a Gold Mirror by an Ultrathin Oxide Nano Lett. 12 2088–94

\bibitem{ref30}
Taflove A, Hagness S C and Piket-May M 2005 Computational Electromagnetics: The Finite-Difference Time-Domain Method The Electrical Engineering Handbook (Elsevier) pp 629–70

\bibitem{ref31}
Thompson C V 2012 Solid-State Dewetting of Thin Films Annu. Rev. Mater. Res. 42 399–434

\bibitem{ref32}
Zhang W, Zheng T, Ai B, Gu P, Guan Y, Wang Y, Zhao Z and Zhang G 2022 Multiple plasmonic hot spots platform: Nanogap coupled gold nanoparticles Applied Surface Science 593 153388

\bibitem{ref33}
Elliott E, Bedingfield K, Huang J, Hu S, De Nijs B, Demetriadou A and Baumberg J J 2022 Fingerprinting the Hidden Facets of Plasmonic Nanocavities ACS Photonics 9 2643–51

\bibitem{ref34}
Noskov A I, Kotlyar A B, Potma E O and Fishman D A Broadband Photo- and Electroluminescence from Bulk Silicon via Strong Photonic Confinement

\bibitem{ref35}
Noskov A I, Kotlyar A B, Katrivas L, Reveguk Z, Garcia E P, Apkarian V A, Galland C, Potma E O and Fishman D A 2026 Overcoming the Indirect Band Gap: Efficient Silicon Emission via Momentum-Engineered Photonic States Nano Lett. 26 5187–94

\bibitem{ref36}
Shoshi J S, Mahdy M R C and Rana M R 2024 On chip all-optical distinguishing of independently placed distinct types of single Rayleigh particle \textit{Heliyon} \textbf{10} e26722

\bibitem{ref37}
Turner W J, Reese W E and Pettit G D 1964 Exciton Absorption and Emission in InP \textit{Phys. Rev.} \textbf{136} A1467–70

\bibitem{ref38}
Wang Q, Taniguchi T, Watanabe K and Smet J H 2025 Boosting the Emission of Momentum Indirect Interlayer Excitons by an Optical Near Field in Misaligned 2D Heterostructures \textit{Nano Lett.} \textbf{25} 14800–7 

\bibitem{ref39}
Shoshi J S, Rodoshi M and Chowdhury M R 2026 Metasurface-driven and polarization controlled broadband on chip tractor beam effect for the cluster of particles \textit{Opt. Express} \textbf{34} 17039

\end{thebibliography}
\end{document}